\documentclass[astrosymb,preprint]{aastex631}

\shorttitle{Sub-pc scale core of FR0s}
\shortauthors{Boughelilba et al.}

\graphicspath{{./}{figures/}}

\begin{document}

\title{On the sub-parsec scale core composition of FR 0 radio galaxies}

\correspondingauthor{Margot Boughelilba}
\email{margot.boughelilba@uibk.ac.at}

\author[ 0000-0003-1046-1647 ]{Margot Boughelilba}
\affiliation{Universit\"at Innsbruck, Institut f\"ur Astro- und Teilchenphysik, 6020 Innsbruck, Austria}
\author[ 0000-0001-8604-7077 ]{Anita Reimer}
\affiliation{Universit\"at Innsbruck, Institut f\"ur Astro- und Teilchenphysik, 6020 Innsbruck, Austria}

\begin{abstract}
Although Fanaroff-Riley (FR) type 0 radio galaxies are known to be the most numerous jet population in the local Universe, they are much less explored than the well-established class of FR I and FR II galaxies due to their intrinsic weakness. Observationally, their nuclear radio, optical and X-ray properties are comparable to the nuclear environment of FR Is. The recent detection of two FR 0s in the high-energy band suggests that like in FR Is, charged particles are accelerated there to energies that enable gamma-ray production. Up to now, only the lack of extended radio emission from FR 0s distinguishes them from FR Is. 
By comparing the spectral energy distribution of FR 0s with that of FR Is and in particular with that of M87 as a well-studied reference source of the FR I population, we find the broadband spectrum of FR 0s exceptionally close to M87's quiet core emission. Relying on that similarity, we apply a lepto-hadronic jet-accretion flow model to FR 0s. This model is able to explain the broadband spectral energy distribution, with parameters close to particle-field equipartition and matching all observational constraints. In this framework, FR 0s are multi-messenger jet sources, with a nature and highly magnetized environment similar to that of the naked quiet core of FR Is.
\end{abstract}


\section{Introduction} \label{sec:intro}

Following the Unified Model for Radio-Loud Active Galactic Nuclei (AGN), radio galaxies have their jets misaligned with the line of sight \citep{Unified_Model_AGN}. For that reason, radio galaxies form the dominant jetted AGN population. Because of this misalignment, the Doppler boosting enhancing the observed flux is small; hence, only a few sources have so far been detected in the gamma-ray band (see, e.g. \citealt{Fermi_AGN, HESS_RG_gamma_ray, MAGIC_RG_gamma_ray}). Blazars, on the other side, with their jets pointing towards Earth, are brighter, but also more rare. 

Based on their extended radio morphology, radio galaxies are usually classified as either faint edge-darkened Fanaroff–Riley type I (FR I) or bright edge-brightened type II (FR II) galaxies. The low-power FR Is are often linked to radiatively inefficient accretion flows, while the more powerful FR IIs are usually associated with more efficient accretion. 
Recently, a new type of radio galaxy has emerged, named FR 0 galaxies  \citep{FR0CAT}. From the radio perspective, FR 0s are similar to FR Is, except for the lack of extended emission (i.e. on a kiloparsec scale). The optical properties of FR 0s are comparable to FR Is, as they are also located in red massive early-type galaxies, and are classified as Low-Excitation Radio Galaxies from a spectroscopic point of view. An X-ray study of a subsample of FR 0s \citep{XrayFR0s} showed that FR 0s have a comparable X-ray luminosity to FR Is in the $2-10 \, \mathrm{keV}$ band, confirming the similarity of the nuclear properties of the two classes. This study also indicates low Eddington-scaled luminosities, hinting towards radiatively inefficient accretion.
In the high-energy domain, the detection of gamma rays from two of them (namely LEDA 55267 and LEDA 58287) has recently been reported \citep{FR0_Paliya} (a third source is mentioned in that paper but it has been removed from the FR0CAT, see \citealt{2019MNRAS.482.2294B}). The stacking analysis of the Fermi-LAT data in \cite{FR0_Paliya} shows that the whole population could be considered as a gamma-ray-emitting class. Previously, \cite{GrandiTol1326} reported the first association of one FR 0, Tol 1326-379, with a gamma-ray source in the Fermi 3FGL catalogue \citep{Fermi3FGL}. The 4FGL source catalogue \citep{Fermi4FGL}, however, reports no gamma-ray counterpart associated with Tol 1326-379, and it is unclear whether this FR 0 is a gamma-ray emitter or not (see, e.g. \citealt{Discussion_TOL1326}).
As of now $\gtrsim 100$ FR 0s (FR0CAT, \citealt{FR0CAT, XrayFR0s}) have been collected, sharing the following properties:  residing at redshift z $\lesssim$ 0.05, the radio sources are located at maximum $2 \arcsec$ from the optical centre, and have a minimum FIRST flux of $5\, \mathrm{mJy}$ at $1.4\, \mathrm{GHz}$. With these properties, FR 0s are shown to be in the order of $\sim 5$ times more numerous than the FR I radio galaxies in the local Universe, which makes them the dominating jet population there \citep{dominantFR0_1,dominantFR0_2}.

Several hypotheses have been proposed so far, to explain the lack of extended radio emission from FR 0s. Evolutionary models consider FR 0s as young sources that evolve into more extended sources. These models are, however, ruled out, due to the distribution of radio sizes in the sample \citep{2019MNRAS.482.2294B}.
Alternatively, \cite{Spin_evolution_Garofalo} discussed the impact of the spin of the SMBH on the power of the associated jets. In this view, FR 0s have been proposed as being driven by a prograde, low-spin SMBH \citep{Garofalo_FR0}, and most of them are not reaching spin values for which non-negligible jets are inferred.

Another approach to gain insight into the true nature of this jet population is linked to their broadband spectral energy distribution (SED). In a recent work, \cite{OurFR0Paper} compiled an average SED of FR 0s to collect information on their radiative environment. Here, we compare for the first time the broadband emission of FR 0s to FR Is, and in particular to M87 as one of the most detailed studied archetypal FR I galaxy.
M87 has been deeply studied, both in its quiet, steady state and in its flaring state. In particular, in 2017, a multi-wavelength campaign focused on the quiet core emission of M87, providing constraints on the core magnetic field, the emission region and the jet properties of M87 \citep{EHTpaper,MagFieldEHT}.

 Section \ref{BB_SED} presents the SED data we collect and discusses the implications taken from the comparison of FR 0s and FR Is. These motivate a model setup for the core region of FR 0s that we describe in Section \ref{Model}. Section \ref{Results} presents the results of our broadband modelling of FR 0s. Our conclusion from this study is discussed in Section \ref{Conclusion}. 

\section{Broadband SED}
\label{BB_SED}
\begin{figure}[h!]
    \centering
    \includegraphics[width = \textwidth]{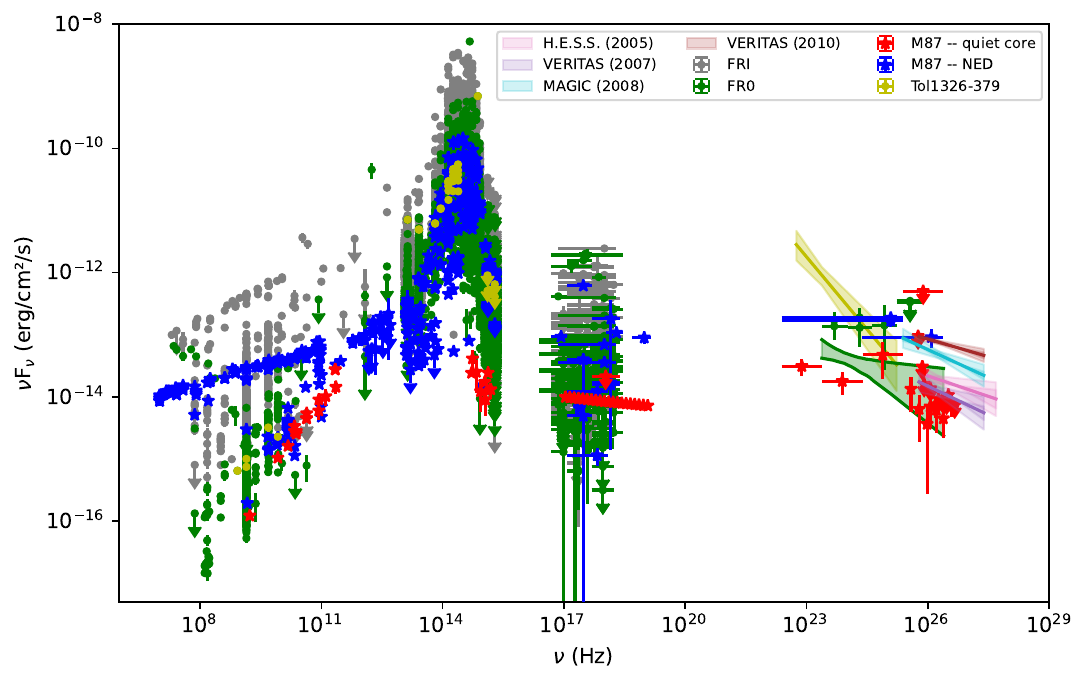}
    \caption{Broadband SED of FR Is (grey dots), FR 0s (green dots and green butterfly), M87 in its 2017 quiet state (red stars). The blue star symbols are the SED of M87 with all the observations available in the NED. The butterfly plots in the very-high-energy gamma-ray range are the power-law spectra fitted to observations of the flaring state of M87 with H.E.S.S in 2005 (pink region), VERITAS in 2007 and 2010 (purple and brown regions respectively) and with MAGIC in 2008 (cyan region). The SED of Tol 1326-379 is shown in yellow, including the butterfly plot at high-energy gamma rays, using the values derived by \cite{GrandiTol1326}. The fluxes from FR Is have been rescaled from their mean distance to the mean distance of FR 0s (i.e. from a luminosity distance $d_\mathrm{L} \sim 1.5 \times 10^{27} \, \mathrm{cm}$ to $d_\mathrm{L} \sim 5.4\times 10^{26} \, \mathrm{cm}$). In the same way, M87 was rescaled to the mean distance of FR 0s.
    The data behind this Figure is available in the online version at the journal webpage. The package contains 3 .fits table files, a python script, and a ReadMe. Included are 114 FR 0, 216 FR I, and one M87 SED tables. The script can be used to read the data files. A list of all the sources is given in the Appendix Table \ref{tab:sources_names}.}
    \label{fig:BB_SED}
\end{figure}

To build the broadband SED of a sample of 114 FR 0s we collected their available data from the \citealt{https://doi.org/10.26132/ned1}\footnote{https://ned.ipac.caltech.edu/}, following the method described in \cite{OurFR0Paper}. 104 sources are taken from the FR0CAT \citep{FR0CAT} (note that 4 of the sources included in the original catalogue have been removed since then, see \cite{2019MNRAS.482.2294B} for more details). The 10 additional sources come from a sample of 19 FR0s studied in the X-ray band \citep{XrayFR0s}, among which 11 were not in the FR0CAT. From these 11 sources, we removed J004150.47-0, which is mentioned to be at the centre of its cluster (Abell85, see \citealt{XrayFR0s}), to avoid flux contamination from the cluster. For these 10  sources, additional observational data from the SSDC SED builder\footnote{https://tools.ssdc.asi.it/SED/} were collected. 
We only use X-ray data if taken with the Neil Gehrels \textit{Swift} Observatory, \textit{XMM}-Newton or \textit{Chandra} telescopes; observations from instruments with a larger angular resolution are discarded to avoid flux contamination from the sources' surroundings. Most Chandra data are taken from the Chandra Point Source Catalog 2.0.1 \footnote{https://cxc.cfa.harvard.edu/csc/} \citep{ChandraCAT} where we use the \textit{flux\_aper90} fluxes (i.e. the reported fluxes represent the background-subtracted fluxes in the modified elliptical aperture). In order to have the most complete data collection available for the two individually gamma-ray detected sources, we built \emph{Swift}-XRT spectra using the online tool\footnote{https://www.swift.ac.uk/user\_objects/}. For LEDA 55267, we used \texttt{XSPEC} \citep{1996ASPC..101...17A} to have a binning of 20 counts per bin to present the data.
The gamma-ray data of LEDA 55267 (SDSS J153016.15+270551.0) and LEDA 58287 (SDSS J162846.13+252940.9) are taken from \cite{FR0_Paliya}. As mentioned in Section \ref{sec:intro}, it is unclear if Tol 1326-379 is a gamma-ray emitter, since no association is reported in the 4FGL catalogue \citep{Fermi4FGL}. Nevertheless, for completeness, the SED of Tol 1326-379 is included into Figure \ref{fig:BB_SED}, with the high-energy emission butterfly representation for an integral photon flux $F_{>1\mathrm{GeV}} = (3.1 \pm 0.8)\times 10^{-10}\, \mathrm{phot}\,\mathrm{cm}^{-2}\,\mathrm{s}^{-1}$ and a spectral index $\Gamma = 2.78 \pm 0.14$, taken from \cite{GrandiTol1326}. Its high-energy slope is much steeper than that of other gamma-ray emitting FR 0s. \cite{Discussion_TOL1326}, however, shows that Tol 1326-379 could be associated with 4FGL J1331.0-3818, with a gamma-flux compatible with the other two gamma-ray-detected sources LEDA 55267 and LEDA 58287, although the association remains ambiguous. For these reasons, we do not consider Tol 1326-379 as a gamma-ray emitting source.

We also collected data of 216 FR I sources listed in the FRICAT \citep{FRICAT} in the same way. Three sources (namely FRICAT 1053+4929, FRICAT 1428+4240, and FRICAT 1518+0613) are also listed as low-luminosity BL Lac Objects \citep{FRICAT, BLLAC} which are therefore not included in our sample.

The multi-wavelength observation of M87's quiet core emission taken in 2017 \citep{EHTpaper} are included, in order to compare the broadband SED of the two classes to a typical FR I source in a quiet state. The flaring states of M87 derived by H.E.S.S in 2005 \citep{HESS_M87_2005}, MAGIC in 2008 \citep{MACIG_M87_2008} and VERITAS in 2007 and 2010 \citep{VERITAS_M87_2007, VERITAS_M87_2010} are also used for the comparison. We use the average fitted values and uncertainties reported all together in \cite{MAGIC_monitoring_M87}.
The list of sources used in this work is shown in Table \ref{tab:sources_names} in the Appendix.

Figure \ref{fig:BB_SED} shows the resulting broadband SEDs of FR 0s as compared to those of M87 and all the other FR Is, all scaled to the mean distance of FR 0s (i.e $z \sim 0.04$). Obviously, FR 0s and FR Is show a very similar spectrum, as expected by the observation in the wavebands discussed in Section \ref{sec:intro}.
The flaring state SED of M87 (that we define as opposed to the quiet state shown in red in Figure \ref{fig:BB_SED} and includes most of the observations shown in blue) is unsurprisingly following the FR Is trend.
The lack of radio emission from FR 0s as compared to FR Is is apparent below $10^{11} \, \mathrm{Hz}$.
What stands out in this comparison is the extreme similarity between M87's quiet core emission and the spectral behaviour of FR 0s, at all wavelengths. 

Contrary to FR 0s, M87 has been extensively studied. Taking the above-highlighted similarity not as a chance coincidence, motivates to apply our knowledge deduced from M87's core observation to gain a deeper understanding -- by modelling -- of the FR 0s. The quiet core study of M87 infers a magnetic field strength of order ($1-30\,\mathrm{G}$) near the core, as well as the presence of an advection-dominated accretion flow \citep{MagFieldEHT}.
Applying such values for the magnetic field to the simplest jet emission model, a one-zone Synchrotron Self-Compton (SSC) model, would result in a synchrotron-dominated SED, with a Compton-dominance ($\nu_\mathrm{comp} L_{\nu,\mathrm{comp}})/(\nu_\mathrm{syn} L_{\nu,\mathrm{syn}}) \ll 1$, where $\nu_\mathrm{syn}$,  $\nu_\mathrm{comp}$ are the synchrotron and Compton peak frequencies respectively, and $L_{\nu,\mathrm{syn}}$, $L_{\nu,\mathrm{comp}}$ the corresponding spectral luminosities at those  peak energies, see \cite{SSCestimates, EHTpaper}. The observed high-energy gamma rays from FR 0s would then have to originate in an emission region further down in the jet. \cite{2019Galax...7...76B}, however, disfavour the large-scale origin of the high-energy radiation from FR 0s.

A model that reproduces the radio-to-gamma-ray quiet core emission of M87 in a one-zone setup was proposed by \cite{Mypaper}. This model focuses on the central region of the AGN, with a jet emission region of a few gravitational radii. Given the compactness of the FR 0s and the SED similarities, we explore here the same type of model for the FR 0 source class. In this model, the high-energy data are explained by the emission of protons, radiating in a high magnetic field. The model also accounts for the accretion flow that is expected in such low-luminosity objects.

\section{Model} \label{Model}

\subsection{Jet} \label{sec:jet}
In this paper, we follow the same approach as in \cite{Mypaper}. 
We consider a continuous cylindrical jet of radius $R'_\mathrm{em}$ and proper length $l' = \Gamma_\mathrm{j} l$, with $l$ being the observed length.
We assume that the emission region contains primary relativistic electrons and protons that are isotropically and homogeneously distributed in the comoving jet frame, and follow a power-law energy spectrum cutting off exponentially, such that the spectral number density $n'_{e,p}(E') \propto E'^{-p_{e,p}}e^{-E'/E'_{\mathrm{max},e,p}}$ cm$^{-3}$, for $E' \ge E'_{\mathrm{min},e,p}$ (where e,p denotes the electrons or the protons, respectively).

The primary particles are continuously injected into the emission region at a rate $q_i$ (cm$^{-3}$s$^{-1}$), where they experience energy losses caused by various interactions. Specifically, we consider photo-meson production, Bethe-Heitler pair production, inverse-Compton scattering, $\gamma$-$\gamma$ pair production, decay of all unstable particles, synchrotron radiation (from electrons and positrons, protons, and $\pi^\pm$, $\mu^\pm$ and $K^\pm$ before their respective decays), and particle escape at a rate $\propto c/R'_\mathrm{em}$. Positrons are treated the same way as electrons. Hence, in the following we will use electrons to refer to the two populations irrespective of their type.

To compute the time-dependent direct emission and cascade component from the jet's particles, we use a particle and radiation transport code (see, e.g. \cite{Anita_matrix_intro}) that is based on the matrix multiplication method described in \cite{Protheroe_Stanev_matrix} and \cite{Protheroe_Johnson_matrix}. The interaction rates and secondary particles' and photons' yields are calculated by Monte Carlo event generator simulations (except for synchrotron radiation, for which they are calculated semi-analytically). These are then used to create transfer matrices, that describe how each particle spectrum will change after a given timestep $\delta t$. To ensure numerical stability, we set $\delta t$ equal to the smallest interaction time for any given simulation. In each timestep, energy conservation is verified. 
The steady-state spectra are calculated by running the simulation until convergence is reached, defined here when $F_\nu(t+\delta t)/F_\nu(t)<1\pm 10^{-3}$.

\subsection{ADAF}
Low-luminosity AGNs are expected to host accretion flows in a radiatively inefficient state. This is characterised by the formation of geometrically thick, optically thin, very hot accretion flows, called Advection-Dominated Accretion Flows (ADAFs, introduced by \citealt{first_adaf_torii, first_adaf} and further developed by e.g. \citealt{Narayan_Yi_original, adafintro}). ADAFs exist only when the accretion rate is sufficiently low ($\dot{M} \lesssim 0.01\dot{M}_\mathrm{Edd}$), and consist of a plasma of thermal electrons and ions, where both components may have different temperatures, $T_e$ and $T_i$ respectively.
Here, we investigate if and how an ADAF component would affect the global SED of FR0s.
We use the ADAF model described in \cite{Mypaper} and will summarize here only the main points. 
In the following, we use the normalized quantities $r=R/R_\mathrm{S}$, with the Schwarzschild's radius $R_\mathrm{S} = 2 r_g = 2.95 \times 10^5 \, m_\mathrm{BH}$, $m_\mathrm{BH}=M_\mathrm{BH}/M_\odot$ and $\dot{m}=\dot{M}/\dot{M}_\mathrm{Edd} = \eta_\mathrm{eff} \dot{M}c^2/L_\mathrm{Edd}$, where $\eta_\mathrm{eff}$ is the radiation efficiency of the standard thin disk ($\eta_\mathrm{eff} \approx 0.1$) and the Eddington luminosity $L_\mathrm{Edd} \simeq 1.3 \times 10^{47} \, m_{\mathrm{BH},9} \, \mathrm{erg}\, \mathrm{s}^{-1}$. 
We obtain the electron temperature by varying $T_e$ using a bisection method to solve the balance equation $ q^{e+} = q^{e-} $ for each radius. Here $q^{e+}$ is the electrons' heating rate, and $q^{e-}$ is their cooling rate. The cooling mechanisms that we consider are synchrotron radiation, bremsstrahlung and Comptonization of the two previous components. The heating mechanisms consist of Coulomb collision between ions and electrons, and viscous energy dissipation.
We make use of the one-zone, height-integrated, self-similar solutions of the slim disc equations derived by \cite{Narayan_Yi_original} to describe the hot plasma. These solutions are appropriate only after the sonic point \citep{Sonic_radius_critical}, corresponding to $\gtrsim 2-5 \, r_g$. The quantities governing the accretion flow depend on the plasma parameter $\beta$, which is the ratio between the gas and the total pressure (i.e., the sum of the magnetic and gas pressure), on the viscosity $\alpha$ and on the heating fraction $\delta_e$ which represents the fraction of viscous energy directly transmitted to the electrons of the plasma. 

Furthermore, we take $\dot{m}$ of the form $\dot{m} = \dot{m}_\mathrm{out} \left(r/r_\mathrm{out}\right)^s$, where $r_\mathrm{out}$ is the outer radius of the ADAF and is associated with an accretion rate $\dot{m}_\mathrm{out}$, and $s$ is a mass-loss parameter (introduced by \citep{Blandford_Begelman_massloss}) that is used to include the presence of outflows or winds from the ADAF.
Upon obtaining the electron temperature, the emitted spectrum from the ADAF is computed, integrating over the radius of the ADAF.

\section{Results} \label{Results}

Motivated by the similarity of the broadband SED of FR 0s to the one of M87's quiet core, we explore parameter sets for the modelling of the FR 0s' emission that are close to the M87 core model of \cite{Mypaper}. 
For the ADAF, we use the same viscosity $\alpha = 0.1$ and heating fraction $\delta_e = 5\times 10^{-3}$. 
We fix the value of the plasma $\beta$ parameter to $\beta = 0.99$, which leads to a magnetic field strength in the central region of the ADAF to be of the order of the estimated jet core magnetic field strength. Lower values of $\beta$ would imply unreasonably large magnetic field strengths. 
For the radial dependence of the accretion rate, parameterized by the index s, we explored values from $0.1$ to $1$ (the larger s is, the more powerful the outflow). Fixing $s$ to $s=0.1$ appears to be a reasonable trade-off between the expected lower power of the jets (compared to M87's jet, where s is set to $s=0.4$) and the radiative flux resulting from such ADAF configurations. We fix $r_\mathrm{out} = 5\times 10^3$, which is a typical value for an ADAF's extension and is well below the size of FR 0s, in the absence of other constraints. 

For a black hole mass range of $10^{7.4} \le M_\mathrm{BH}/M_\odot \le 10^{9}$ \citep{FR0CAT} (with a mean value of $M_\mathrm{BH} \approx 10^{8.4} \, M_\odot$) for the FR 0 source class, one expects a lower ADAF X-ray luminosity than for the FR Is possessing black holes with a mass range of $10^{8} \le M_\mathrm{BH}/M_\odot \le 10^{9.5}$ (and with a mean value of $M_\mathrm{BH} \approx 10^{8.55} \, M_\odot$).

For adjusting the accretion rate in order to match the observations, we follow a step-by-step procedure. First, the accretion rate is set to the highest allowed value (for a given $\alpha$, $\beta$ and $M_\mathrm{BH}$, \citealt{Narayan_Yi_original}). Then, we compute the associated magnetic field in the central region (namely where $R \le R'_\mathrm{em}$). If the magnetic field strength in the ADAF there exceeds the value of the jet core magnetic field, we decrease the accretion rate accordingly to reach this value. The accretion rate can be further reduced if needed to match the observations. 
The ADAF spectrum is then calculated with the method described above. We do so for the two gamma-ray detected sources, as well as for the 23 other sources where X-ray data are available.
The resulting SED is a combination of the ADAF component, the jet component and the host galaxy's modified blackbody.

FR 0s' jets are expected to be less powerful than FR Is' and only mildly relativistic \citep{jets_fr0}. Therefore, we explored parameters similar to those used to model M87 \citep{Mypaper} except a lower value for the average relative jet bulk velocity $\beta_\mathrm{j}$, namely $\beta_\mathrm{j} = 0.55$, and a jet inclination with respect to the line of sight of $20^\circ$. 
We consider a magnetic field strength in the range $\sim 10-50 \, \mathrm{G} $, primary particle spectral indices of $1.7-2.3$ and an emission region of a few to hundreds of gravitational radii in size. Lower values for the magnetic field strength imply X-ray fluxes that do not reach the observed level: For the same ADAF parameters, lowering the magnetic field strength implies decreasing the accretion rate which results in correspondingly lower X-ray luminosities. Satisfactory results are obtained when using magnetic field strengths in the range $25-50\,\mathrm{G}$. The emission region's size varies from $R'_\mathrm{em} =  4 \times 10^{15} \, \mathrm{cm}$ for $B = 25 \, \mathrm{G}$ to $R'_\mathrm{em} =  1.2 \times 10^{15} \, \mathrm{cm}$ for $B = 50 \, \mathrm{G}$, in order not to overshoot the available jet power (predicted in the range $10^{42.5}-10^{43.5}\, \mathrm{erg}\,\mathrm{s}^{-1}$ for FR0s (\citealt{OurFR0Paper, Estimate_Jet_power}).

To allow the jet emission to reach the X-ray energies and corresponding flux levels, a hard slope is preferred and better fits are achieved with an electron spectral index of $p_\mathrm{e} = 1.7$. 
The proton spectral index is mainly constrained by the resulting jet power. For that reason, we keep models with $p_\mathrm{p} = 1.7$. The maximum proton energy varies from $E'_{\mathrm{max},p} = 10^9 \, \mathrm{GeV}$ to $E'_{\mathrm{max},p} = 5.5\times 10^9 \, \mathrm{GeV}$. We model the two gamma-ray detected sources individually, for the subthreshold sample we aim at an average description of the population. The injection parameters and some resulting quantities for the different models are given in Tables \ref{table:parameters_25G} and \ref{table:parameters_50G}.
\begin{table}
\caption{Jet parameters used in the case $B = 25\,\mathrm{G}$. The size of the emission region is $R'_\mathrm{em}$, $n'_\mathrm{inj,e(p)}$ is the electron (proton) number density injection rate, and both types of particles are injected with spectral indices $p_\mathrm{e,p} = 1.7$, following the spectral shape described in \ref{sec:jet}. $u'_\mathrm{part,ss}/u'_\mathrm{B}$ and $L'_\mathrm{jet,ss}$  represent the energy density ratio and jet power respectively, after the steady state is reached in the emission region.}
\begin{center}
\begin{tabular}{|m{85pt}||m{60pt}|m{60pt}|m{60pt}|}
    \hline
    & LEDA 55267 & LEDA 58287 & Subthreshold sample \\
    \hline
    $R'_\mathrm{em}$ (cm) & $4.0 \times10^{15}$ & $4.0 \times10^{15}$ & $4.0 \times10^{15}$\\ 
    $n'_\mathrm{inj,e}$ (cm$^{-3}\,$s$^{-1}$) & $1.4 \times 10 ^{-2}$ & $1.6 \times 10 ^{-3}$ & $4.8 \times 10 ^{-4}$ \\ 
    $n'_\mathrm{inj,p}$ (cm$^{-3}\,$s$^{-1}$) & $3.9 \times 10 ^{-6}$ & $3.7 \times 10 ^{-6}$ & $1.8 \times 10 ^{-6}$ \\ 
    $E'_\mathrm{min,e}$ (MeV) & $0.5$ & $0.5$ & $0.5$ \\ 
    $E'_\mathrm{max,e}$ (MeV) & $1.2 \times 10^4$ & $8.0 \times 10^3$ & $1.5 \times 10^4$ \\ 
    $E'_\mathrm{min,p}$ (GeV) & $1.0 $  & $1.0 $ & $1.0 $ \\ 
    $E'_\mathrm{max,p}$ (GeV) & $3.0 \times 10 ^{9}$ & $5.5 \times 10 ^{9}$ & $2.0 \times 10 ^{9}$ \\ 
    $p_\mathrm{e} = p_\mathrm{p}$ & $1.7$ & $1.7$ & $1.7$ \\ 
    $u'_\mathrm{part,ss}/u'_\mathrm{B}$ & $2.1 \times 10^{-1}$ & $9.7 \times 10^{-2}$ & $3.5 \times 10^{-2}$\\
    $L'_\mathrm{jet,ss}$  ($\mathrm{erg}\,\mathrm{s}^{-1}$) & $3.6 \times 10^{43}$ & $3.3 \times 10^{43}$ & $3.1 \times 10^{43}$\\
    \hline
\end{tabular}
\end{center}
\label{table:parameters_25G}
\end{table}

\begin{table}
\caption{Same as in table \ref{table:parameters_25G} for the case $B = 50\,\mathrm{G}$.}
\begin{center}
\begin{tabular}{|m{85pt}||m{60pt}|m{60pt}|m{60pt}|}
    \hline
    & LEDA 55267 & LEDA 58287 & Subthreshold sample \\
    \hline
    $R'_\mathrm{em}$ (cm) & $1.2 \times10^{15}$ & $1.2 \times10^{15}$ & $1.2 \times10^{15}$\\
    $n'_\mathrm{inj,e}$ (cm$^{-3}\,$s$^{-1}$) & $4.2 \times 10 ^{-1}$ & $3.2 \times 10 ^{-2}$ & $1.9 \times 10 ^{-2}$ \\
    $n'_\mathrm{inj,p}$ (cm$^{-3}\,$s$^{-1}$) & $1.3 \times 10 ^{-4}$ & $1.9 \times 10 ^{-4}$ & $8.2 \times 10 ^{-5}$ \\
    $E'_\mathrm{min,e}$ (MeV) & $0.5$ & $0.5$ & $0.5$ \\
    $E'_\mathrm{max,e}$ (MeV) & $8.0 \times 10^3$  & $8.0 \times 10^3$ & $8.0 \times 10^3$  \\
    $E'_\mathrm{min,p}$ (GeV) & $1.0$  & $1.0$ & $1.0$ \\
    $E'_\mathrm{max,p}$ (GeV) & $3.0 \times 10 ^{9}$ & $4.0   \times 10 ^{9}$ & $1.5 \times 10 ^{9}$ \\
    $p_\mathrm{e} = p_\mathrm{p}$ & $1.7$ & $1.7$ & $1.7$ \\
    $u'_\mathrm{part,ss}/u'_\mathrm{B}$ & $4.8 \times 10^{-1}$ & $2.9 \times 10^{-1}$ & $1 \times 10^{-1}$\\
    $L'_\mathrm{jet,ss}$  ($\mathrm{erg}\,\mathrm{s}^{-1}$) & $1.5 \times 10^{43}$ & $1.3 \times 10^{43}$ & $1.1 \times 10^{43}$\\
    \hline
\end{tabular}
\end{center}
\label{table:parameters_50G}
\end{table}

Our best-fit accretion rate values depend on the magnetic field strength present in this region. We find values for the accretion rate at the outer boundary of the flow of $\dot{m}(r = r_\mathrm{out}) \sim 6\times 10^{-4}-  2\times 10^{-3}$ when the jet's magnetic field strength is $B = 25 \, \mathrm{G}$ whereas $\dot{m}(r = r_\mathrm{out}) \sim 1\times 10^{-3} - 4\times 10^{-3}$ for $B = 50 \, \mathrm{G}$.

\begin{figure}[h!]
    \centering
    \includegraphics[width = \textwidth]{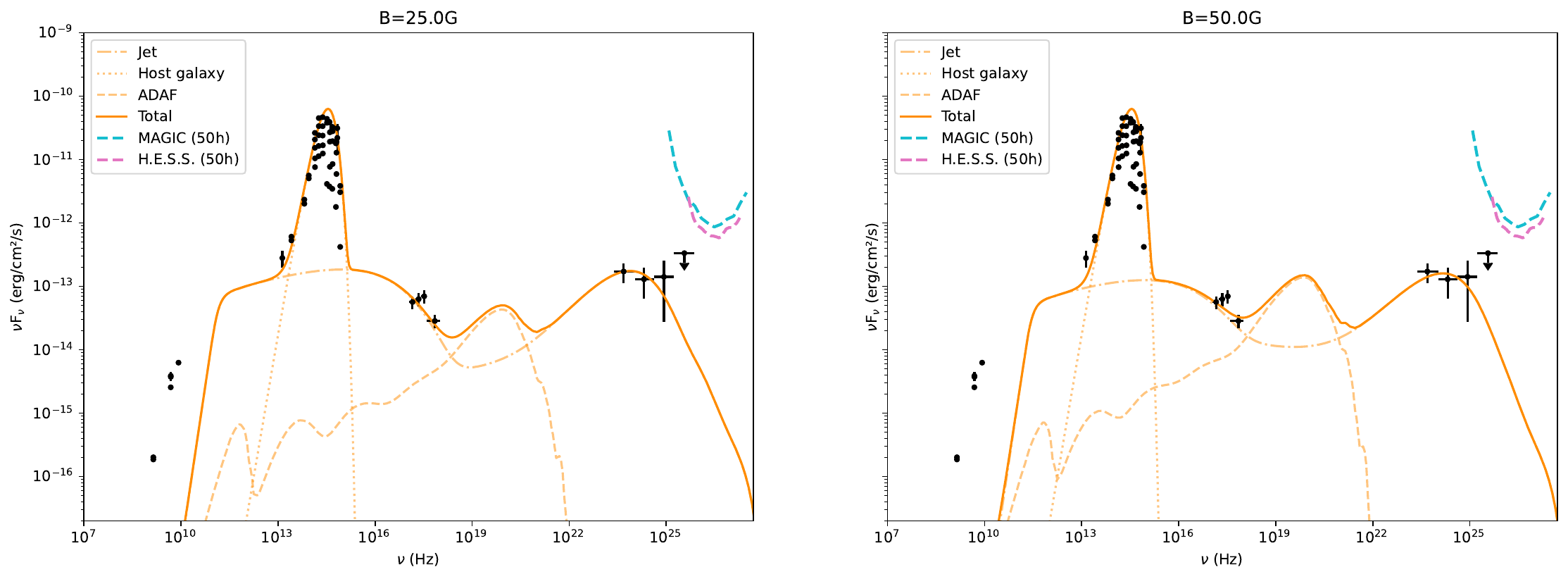}
    \caption{SEDs of LEDA 55267. The dotted line is the modified blackbody modelling the host galaxy emission, the dashed line is the emission coming from the ADAF, the dash-dotted line is the total jet's emission and the solid line represents the total emission of the source and is the sum of the three components. The differential fluxes sensitivities for 50 hours of observation with MAGIC \citep{MAGICsensitivity} and H.E.S.S \citep{HESSsensitivity} are shown with the cyan and pink dashed lines respectively. \textit{Left:} for a magnetic field strength of 25G in the jet. \textit{Right:} same for a magnetic field strength of 50G.} 
    \label{fig:LEDA55267}
\end{figure}

In Figure \ref{fig:LEDA55267}, we present the SED  of LEDA 55267 and its model representations for a jet magnetic field strength of 25G and 50G, from left to right, respectively. 

\begin{figure}[h!]
    \centering
    \includegraphics[width = \textwidth]{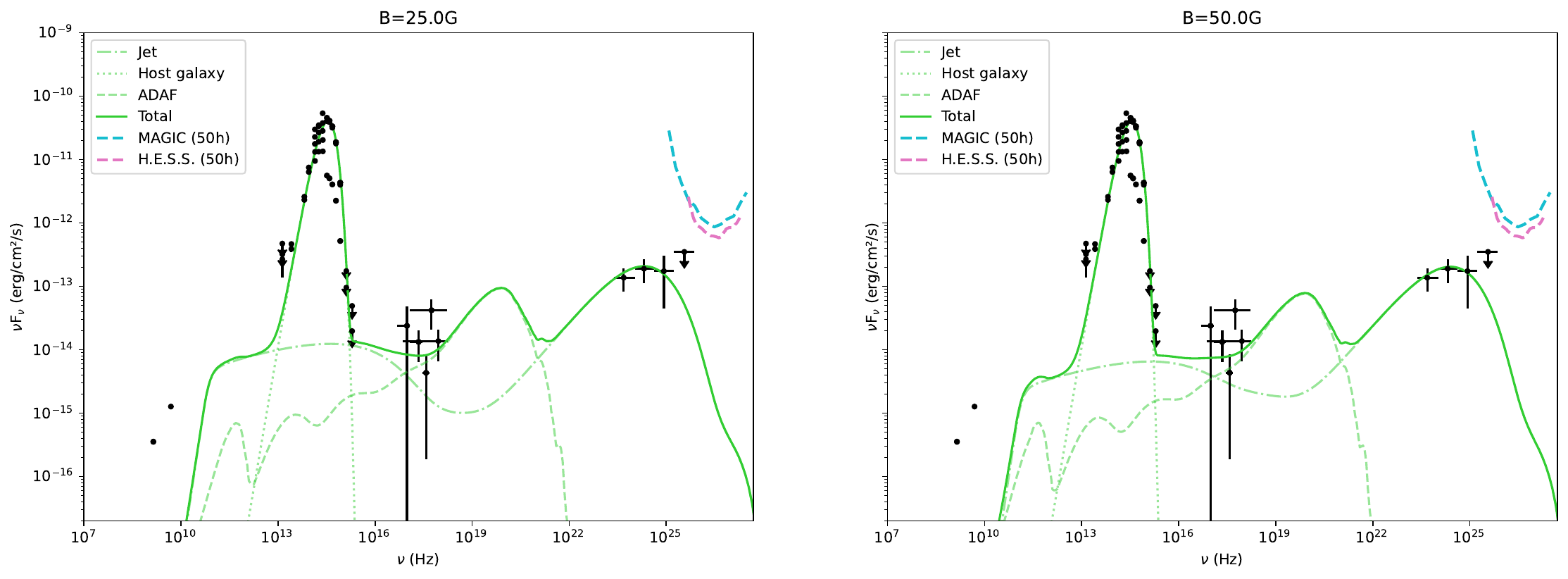}
    \caption{Same as Figure \ref{fig:LEDA55267} but for LEDA 58287} 
    \label{fig:LEDA58287}
\end{figure}

The same is shown in Figure \ref{fig:LEDA58287} for the second gamma-ray detected source, namely LEDA 58287.

\begin{figure}[h!]
    \centering
    \includegraphics[width = \textwidth]{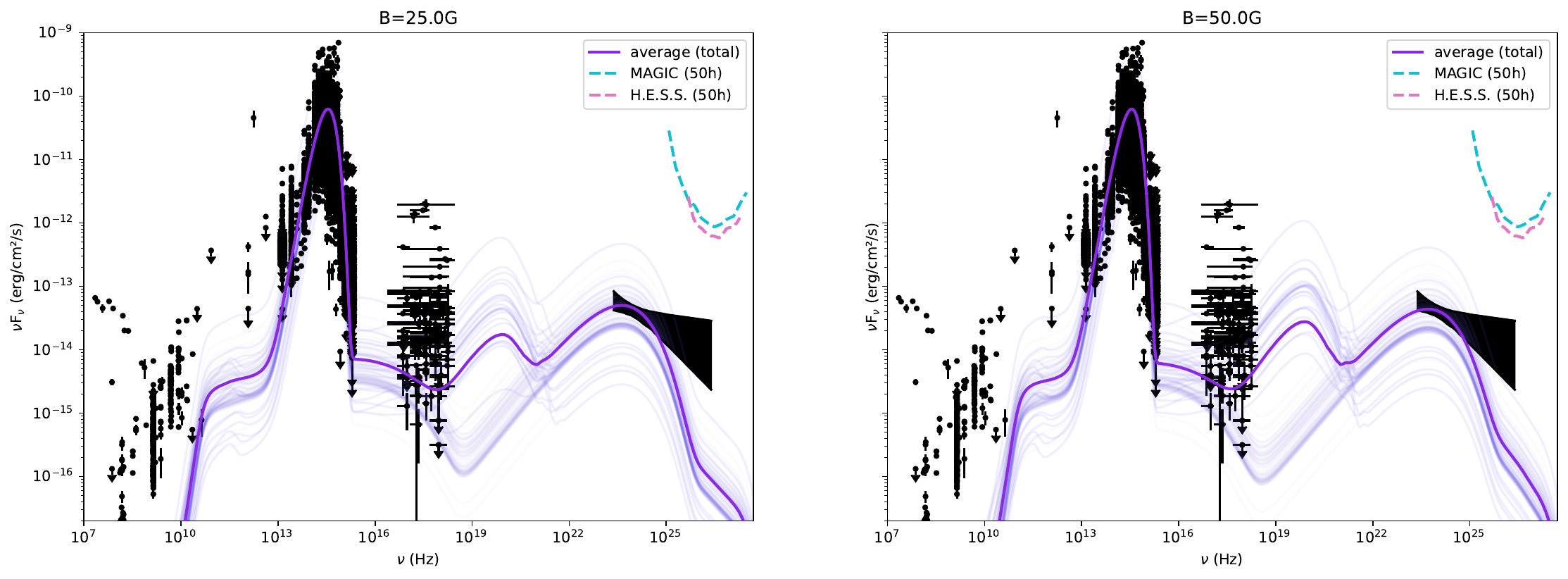}
    \caption{SEDs of the 112 sources that are not individually detected in the gamma-ray band. The faint purple lines are the individual fluxes of the 112 sources (see main text for details) and the purple blue line is the average of the 112 models. The differential fluxes sensitivities for 50 hours of observation with MAGIC, H.E.S.S are shown with the cyan and pink dashed lines respectively. \textit{Left:} for a magnetic field strength of 25G in the jet. \textit{Right:} same for a magnetic field strength of 50G.}
    \label{fig:sub}
\end{figure}

As described above, the 23 subthreshold sources with X-ray data possess a modelled ADAF and a jet. All the 112 subthreshold sources are modelled with the same jet parameters. For each source, the observed flux is calculated from the emitted luminosity, given their respective distance to Earth. The corresponding SEDs are displayed in Figures \ref{fig:sub}, in faint purple, for two magnetic field strengths, 25G on the left panel and 50G on the right panel respectively. 

The average SED of the 112 FR 0s is shown as a solid plain blue line there. The TeV flux predicted by our models lies far below the sensitivity curves of the current Cherenkov telescopes\footnote{The MAGIC differential sensitivity is available in machine-readable format at: \url{https://magic.mpp.mpg.de/newcomers/magic-team/technical-implementation0/} and the H.E.S.S curve is adapted from \cite{HESSsensitivity}, see \url{https://www.cta-observatory.org/science/ctao-performance/}}.

We predict a strong MeV contribution from the ADAF to the overall sources' SED (even if slightly less important in the case of $B = 25 \, \mathrm{G}$). This component could be probed by future MeV gamma-ray instruments like e-ASTROGAM \citep{eASTROGAM} or the All-sky Medium Energy Gamma-ray Observatory eXplorer (AMEGO-X) \citep{AMEGO_Mission, AMEGO_MeV}.. 

The steady-state jet power is estimated by $L'_\mathrm{jet,ss} = \pi R^{'2}_\mathrm{em} \Gamma_\mathrm{j}^2 \beta_\mathrm{j} c  \sum_{i}^{}u'_i$ where $u'_\mathrm{i}$ is the energy density of radiation, electrons, protons ($u'_\mathrm{part,ss}$) and magnetic field ($u'_\mathrm{B}$) respectively. We assume a neutral jet and hence account for cold protons to balance the electrical charge. In the case of $B = 25 \, \mathrm{G}$, we find the jet to be slightly magnetically dominated, i.e. $u'_\mathrm{part,ss}/u'_\mathrm{B} \approx 3.5\times 10^{-2} - 0.1$. For $B = 50 \, \mathrm{G}$, the jet composition is very close to equipartition, i.e. $u'_\mathrm{part,ss}/u'_\mathrm{B} \approx 0.1 - 0.5$. The resulting jet power is in the range $(1.1 - 1.5) \times 10^{43} \, \mathrm{erg}\,\mathrm{s}^{-1}$ for $B = 50 \, \mathrm{G}$ and $(3.0 - 3.6) \times 10^{43} \, \mathrm{erg}\,\mathrm{s}^{-1}$ for $B = 25 \, \mathrm{G}$.
Our calculated neutrino output of the models predicts neutrino fluxes far below the current instruments' sensitivities (peak fluxes lie at $\lesssim 10^{-13} \, \mathrm{GeV}\,\mathrm{cm}^{-2}\,\mathrm{s}^{-1}$ with a peak energy of $E_\mathrm{peak} \sim 10^{17-18} \, \mathrm{eV}$). 

\section{Conclusion} \label{Conclusion}

Aiming to gain a deeper understanding of the dominating jet population in the local Universe, Fanaroff-Riley type 0 radio galaxies, we compared these to the more extended but comprehensively studied FR Is. 
We found that the broadband SED of FR 0s is extremely similar to the archetypal FR I, M87, during its quiet steady state (described in detail in \citealt{EHTpaper}). The similarity goes from the core radio emission to the X-ray band, and up to gamma-rays for two individual sources detected in the high-energy band.

This motivates to consider an environment described by physical parameter values that is comparable to M87's quiet core. To test this, we applied a one-zone lepto-hadronic jet model, combined with the emission of an advection-dominated accretion flow to the FR 0 population. Alternatively, two-zones models, like a spine-sheath jet structure, are not rejected. Indeed, recently, \cite{pc-scale-fr0,emerlin_fr0,jets_fr0} showed that FR 0s have a smaller jet-to-counterjet ratio than FR Is, on pc-scale. This suggests that FR 0s' jets are mildly, or even not, relativistic, which can also be interpreted as the presence of a faint relativistic spine and a dominant slow sheath structure in the jet. In this framework, if FR 0s' jets are seen at a large viewing angle, as indicated by observations, mainly the sheath emission would be observed, and our results can be interpreted as the emission from this zone, at the first order.
In the one-zone model context, we found that a compact subparsec-scale jet-flow emission region (from a few to a thousand gravitational radii for the jet, to $5\times 10^3 \, r_g$  for the ADAF, leading to a global region size of $\sim 6\times 10^{-3} -  0.3 \, \mathrm{pc}$) is able to explain the nuclear multiwavelength SED of FR 0s, provided that a magnetic field strength of $25-50\,\mathrm{G}$ is reached in the core region. As reviewed by \cite{review_fr0}, lower values of the magnetic field strength are expected to prevent the formation of large-scale jets and explain the lack of extended emission in FR 0s. \cite{Nikita_FR0} explore broadband modelling scenarios with such low field strengths, where then the jet's composition is strongly particle-dominated, and leptons can account for the high-energy observations.

In this model, the jet of FR 0s is mildly relativistic, with a velocity $\beta_\mathrm{j}c = 0.55\,c$, which is consistent with the value obtained by \cite{jets_fr0} when observing the core of FR 0s in comparison to FR Is. The jet contributes mainly to the radio and gamma-ray band. The optical observations are dominated by the host galaxy. The jet and the ADAF both contribute to the X-ray band, predicting a strong ADAF-dominated MeV flux component.

As protons are, in this framework, accelerated up to $\sim 6 \times 10^{18} \, \mathrm{eV}$, FR 0s are multi-messenger sources and could contribute to the cosmic-ray flux up to the ankle ($E' \approx 10^{18} \, \mathrm{eV}$, see also \citealt{OurFR0Paper,ICRC_JonPaul}).

In this view, we find that FR 0s, given their observed nuclear properties and their broadband SED, are of a similar nature as that of the naked quiet core of FR Is, whose best-studied representation is the quiet core of M87.

\begin{acknowledgments}
This work acknowledges financial support from the Austrian Science Fund (FWF) under grant agreement
number I 4144-N27. MB has for this project received funding from the European Union’s Horizon 2020 research and innovation program under the Marie Sklodowska-Curie grant agreement No 847476. The views and opinions expressed herein do not necessarily reflect those of the European Commission. MB wishes to thank Paolo Da Vela and Giacomo Bonnoli for the fruitful discussions and insightful comments on this paper. 

\end{acknowledgments}

\software{This work benefited from the following software: NumPy \citep{numpy}, Matplotlib \citep{matplotlib}, pandas \citep{pandas, panda_software}, jupyter notebooks \citep{ipython}.}

\bibliography{references}{}
\bibliographystyle{aasjournal}

\appendix
\begin{deluxetable}{c}[h!]
\tablecaption{The FR 0 and FR 1 source list\label{tab:sources_names}}
\tablehead{
\colhead{Name}
}
\startdata
\cutinhead{FR 0}
SDSS J010101.12-002444.4 \\
SDSS J010852.48-003919.4 \\
SDSS J011204.61-001442.4 \\
\cutinhead{FR I} 
SDSS J002900.90-011341.7 \\
SDSS J003930.52-103218.6 \\
SDSS J004148.22-091703.1 \\
\enddata
\tablecomments{Table \ref{tab:sources_names} is published in its entirety in the electronic edition of the {\it Astrophysical Journal Letters}.  A portion is shown here for guidance regarding its form and content.}
\end{deluxetable}

\end{document}